# The Removal of Single Layers from Multi-Layer Graphene by Low Energy Electron Stimulation

*Jason D. Jones, Rakesh K. Shah, Guido F. Verbeck, and Jose M. Perez\**

[*]  Prof. J. M. Perez
Department of Physics
University of North Texas
Denton, TX 76203 (USA)
E-mail: jperez@unt.edu

   J. D. Jones, R. K. Shah
Department of Physics
University of North Texas
Denton, TX 76203 (USA)

   Prof. G. F. Verbeck
Department of Chemistry
University of North Texas
Denton, TX 76203 (USA)

Keywords:  thinning, plasma, electron irradiation, graphene, Raman spectroscopy

The removal of single atomic layers from multi-layer graphene using a He plasma is reported. By applying sample biases of -60 and +60 V during He plasma exposure, layer removal is found to be due to electrons instead of He ions or neutrals in the plasma. The rate of layer removal depends on exposure time, sample bias and pre-annealing treatments. Optical contrast microscopy and atomic force microscopy studies show that the removal of C atoms occurs approximately one layer at a time across the entire multi-layer sample with no observable production of large pits or reduction in lateral dimensions. Layer removal is proposed to arise from the electron-stimulated dissociation of C atoms from the basal plane. This process differs from plasma techniques that use reactive species to etch multi-layer graphene.

# 1. Introduction

Graphene and multi-layer graphene have recently received considerable interest due to their potential applications in electronic and mechanical devices.[1,2] Since the properties of multi-layer graphene–such as electrical conductivity, optical transparency and mechanical strength–depend on the number of layers, $n$, methods of reducing $n$ in regions of multi-layer graphene may have useful applications. For example, in flat panel displays and solar cells, thin regions are required for optical transparency while thicker, more opaque regions–having higher electrical conductivity–are desired for use as interconnects.[3-5] In this paper, we demonstrate a simple method for thinning exfoliated multi-layer graphene approximately one layer at a time using low-energy electron irradiation from a He plasma. By studying the effects of plasma exposure at sample biases of -60 and 60 V, we find that electrons rather than He ions or neutrals in the plasma are responsible for thinning. This technique removes approximately one layer at a time across the entire multi-layer sample without producing large pits or a reduction in lateral dimensions. We propose that the removal of layers is due to electron stimulated dissociation of C atoms from the basal plane of graphene, similar to the phenomena of electron stimulated desorption of adsorbates on surfaces and fragmentation of molecules in gases.[6,7] Since plasmas similar to those in our experiments are compatible with microfabrication technology, this simple technique may be a scalable means of thinning selected regions of multi-layer graphenes for device applications.

Other thinning techniques typically use reactive atomic species such as O or H in gases or plasmas to etch the basal plane of graphene and can produce deep multi-layer pits and reduction in lateral dimensions of the sample.[8-13] Recently, laser heating and acid treatment techniques have been reported to thin multi-layer graphene.[14,15] In the former, a scanned laser beam from a Raman system is used to heat and remove top layers by oxidation. This approach is limited dimensionally due to the slow scan rate of the laser and may not be feasible for large-scale patterning. In addition, the number of layers removed cannot be controlled since all the layers except the one closest to the substrate are removed. In the latter technique, Zn is sputtered onto the sample, and the resulting impact damage to the top layer allows the layer to be etched using acid. This method requires repeated sputtering and acid treatment to remove successive layers, and, consequently, may not be feasible for removing many layers. In addition, the removal of layers from as created chemical vapor deposition (CVD) bi- and tri-layer films by the chemisorbtion of atomic oxygen followed by the bombardment of argon atoms has been recently reported.[16] Each layer removed using this technique requires successive oxygen and argon treatments, and thus may not be feasible for removing many layers. We note the high defect density of such films thinned using this technique, even after annealing at 1000 °C. Furthermore, the use of heavy ions rather than electrons in their complex process could potentially lead to highly defective films. Other techniques implement the use of a H-N plasma to thin reduced multi-layer graphene oxide at sample temperatures of 400 °C.[17] However, the important role of electrons in the process was not addressed in this work. Furthermore, reduced graphene oxide samples typically have high defect densities limiting their device applications.

## 2. Results and Discussion

We use optical contrast microscopy for locating flakes produced by exfoliation. Subsequent micro-Raman spectroscopy is carried out for determining $n$. Inspection of the 30 cm$^{-1}$ full-width-at-half-maximum of the 2D peak at about 2700 cm$^{-1}$ is sufficient for mono-layer graphene (MLG) identification; this value doubles for bi-layer graphene (BLG) and multi-layer graphene.[18] A linear relationship has been reported between $I_G/I_{Si}$ and $n$, where $I_G$ and $I_{Si}$ are the integrated intensities of the $G$ peak at 1580 cm$^{-1}$ and the Si peak at 520 cm$^{-1}$, respectively.[19] From this relationship, $n$ for thicker layer samples can be extrapolated.[19]

We use a capacitively coupled RF plasma system operating at 19-21 MHz and 50 W. A detailed description of the plasma system is given elsewhere.[20] Plasma exposures are carried out within a stainless steel chamber. With the sample floating, there is a voltage of about 25 V between the sample and chamber due to plasma sheaths.[20] An overwhelming majority of the species in a He plasma consists of He atoms, metastable helium atoms ($He^*$), helium ions ($He^+$), dimer helium metastable atoms ($He_2^*$), dimer helium metastable ions ($He_2^+$), and electrons.[21-23] A DC bias may be introduced between the sample and chamber during plasma exposure using an external power supply. A positive bias on the sample attracts electrons, repels positive ions, and leaves neutral species unhindered from diffusing to the surface. A negative bias repels electrons, attracts positive ions and also leaves neutrals unhindered. A marked effect on the surface as a result of only positive bias can then be attributed to the action of electrons, while an effect only under negative bias to positive ions. If neutrals play a role in an effect, then their contribution would be observable under both of these biasing conditions.

**Figure 1** (a) shows an optical image of a graphite flake produced by exfoliation prior to He plasma exposure. Figure 1 (b) is an optical image of the same flake after exposure to He plasma for 13 min with the sample electrically floating. We observe an overall reduction in optical contrast for all regions of the flake. The MLG region indicated in Figure 1 (a) has completely disappeared while the BLG region is still observable but appears to have the contrast of MLG. Figures 1 (c) and (d) are atomic force microscopy (AFM) images of the boxed regions indicated in Figures 1 (a) and (b), respectively. No reduction of lateral dimensions is observed for any region of the flake after thinning. **Figure 2** (a) shows step height profiles between the Si substrate and MLG region before and after He plasma exposure. After exposure, the step height is reduced from approximately 6 Å to zero, consistent with the complete removal of the MLG. Figure 2 (b) shows a step height analysis for the BLG region, showing that the step height is reduced by a factor of two from approximately 13 to 6 Å, consistent with the removal of one layer. The analysis was done using the program WSxM v5.0.[24] In Figure 2 (c), we compare $I_G/I_{Si}$ versus $n$, for $n$ = 1, 2, 3, 4, 5, before and after plasma exposure. Before exposure, the flake regions exhibit a linear relation in $I_G/I_{Si}$ vs. $n$ with a slope of approximately 0.11 $\Delta(I_G/I_{Si})$/layer. After exposure, the value of $I_G/I_{Si}$ for each region having $n$ = 2, 3, 4, 5 decreases by approximately 0.13. This decrease is comparable to the slope of 0.11 $\Delta(I_G/I_{Si})$/layer, indicating that one layer has been removed from each region. For the MLG region, no Raman $G$ peak is observable after exposure. In addition, after exposure, the $I_G/I_{Si}$ vs. $n$ plot remains linear with a slope of approximately 0.11 $\Delta(I_G/I_{Si})$/layer. These results corroborate the removal of one layer from all the regions. A residual depression in the shape of the former MLG region is present in the AFM image in Figure 1 (d), as indicated by the arrow. We do not believe that this depression is due to residual carbon since it registers as a depression instead of an elevation in the AFM image, and the Raman signal from this region did not show $G$ or $2D$ peaks. It is possible that during the removal of the final layer in contact with the substrate the graphene interacts with the $SiO_2$, causing a stronger interaction between the AFM probe and the sample.

**Figures 3** (a) - (c) show optical images of a flake before and after two consecutive He plasma exposures, denoted by numerals I, II and III, respectively. Figure 3 (a) shows an optical image of the flake prior to exposure for 40 min with the sample electrically floating. After the first exposure, an overall reduction in optical contrast equivalent to the removal of one atomic layer is observed for all layers of the flake, as shown in Figure 3 (b). The MLG region in Figure 3 (a) has completely disappeared while the BLG and multi-layer regions remain optically visible. Figure 3 (c) shows an optical image of the same flake after a second He plasma exposure for 10 min. A further reduction of optical contrast corresponding to the removal of two atomic layers is observed for all layers. In Figure 3 (d), we compare $I_G/I_{Si}$ versus $n$, for $n$ = 1, 2, 6, and 8, before and after the two subsequent He plasma exposures. Before exposure, the indicated flake regions exhibit a linear relation in $I_G/I_{Si}$ vs. $n$ with a slope of approximately $0.11\Delta(I_G/I_{Si})$/layer. After the first exposure, the value of $I_G/I_{Si}$ for each region originally having $n$ = 2, 6, and 8 decreases by approximately 0.09, as indicated by the squares in Figure 3 (d). This decrease is comparable to the slope of $0.11\Delta(I_G/I_{Si})$/layer, indicating that one layer has been removed from each region. After the second exposure, the value of $I_G/I_{Si}$ for each region originally having $n$ = 6 and 8 decreases by approximately 0.18, as indicated by the triangles in Figure 3 (d). This decrease is comparable to twice the slope of $0.11\Delta(I_G/I_{Si})$/layer, indicating that two layers have been removed from each region. Figure 3 (e) shows the evolution of the Raman spectra for the $n$ = 1, 2, 6 and 8 regions before and after the two consecutive He plasma exposures. For the MLG region, the Raman signal disappears after the first exposure. For the BLG region, a $D$ peak with $I_D/I_G$ = 1.24, where $I_D$ is the integrated intensity of the $D$ peak, appears after the first exposure, indicating the introduction of defects. Values of $I_D/I_G$ for different BLG regions thinned to MLG under the same plasma conditions ranged from approximately 0.65 to 2.1. The lower values of $I_D/I_G$ in this range are comparable to those reported for MLG thinned from BLG using laser heating and acid treatment.[14,15] As discussed below, the $D$ peak we observe may be due, in part, to incomplete removal of the top layer. As the defect peaks reflect the formation of $sp^3$ in the graphene, the edge and surface carbons are likely bound to hydrogen. This hydrogen can easily be found in the presence of low energy electrons and plasma due to the $H_2O$ presence on the surfaces.[25,26] We note that the previously reported BLG region thinned to MLG using acid treatment was not as-grown BLG, but BLG created by transferring a MLG film onto another MLG film.[15] Such a structure has a greater inter-layer spacing than BLG that is produced by exfoliation in which the layers are bonded by van der Waals forces. Due to the increased inter-layer distance, the previously reported as-created BLG structure may have a bottom layer that is not as susceptible to defect formation during removal of the top layer. As shown in Figure 3 (e), the $n = 6$ region has $I_D/I_G = 1.04$ and 1.32 after the first and second exposures, respectively; while the $n = 8$ region has $I_D/I_G = 0.74$ and 0.88 after the first and second exposures, respectively. Thus, after the first exposure, the multi-layer regions exhibit smaller values of $I_D/I_G$, consistent with defects occurring predominately on the top layers of the sample. Although the existence of defects in the thinned layers may limit their applications in high-speed electronic devices, the increase in optical transmittance may be useful for thin film transparent conducting electrodes.

**Figures 4** (a) and (b) show optical images of a flake before and after He plasma exposure for 30 min at a sample bias of -60 V. After exposure, no reduction of optical contrast is observed for the MLG, BLG and multi-layer regions, indicating that no layers have been removed. Figure 4 (c) shows an optical image of a flake prior to plasma exposure for 30 min at a bias of +60 V. Electrons in the plasma typically have kinetic energies of about 2 eV.[20] Thus, a positive bias of $V$ volts on the sample results in incident electrons having energies of $V \pm 2$ eV. The flake in Figure 4 (c) contained MLG, BLG and multi-layer regions up to about $n = 30$. After exposure, the entire flake disappeared, indicating an average removal rate of at least 1 layer/min. We conclude that electrons, rather than positive ions or neutrals, are responsible for layer removal. We exposed samples to plasma under +60 V bias for shorter periods of time to determine the rate of layer removal. Figures 4 (d) and (e) show optical images of a flake before and after plasma exposure for 5 min at a bias of +60 V. As shown, the MLG and BLG regions have been completely removed while the remaining TLG region is still visible, indicating that approximately two layers have been removed at an average rate of 0.4 layers/min. This rate is a factor of two less than the minimum rate for the sample in Figure 4(c) that was exposed at +60 V for 30 min. Thus, the removal rate is lower at the beginning of exposure. We attribute the increase in removal rate for longer exposures to electron interaction with underlying layers created by top-layer vacancies. Therefore, simultaneous removal of more than one layer would speed up the thinning process.

We investigated the effects of pre-annealing samples on layer removal. Samples were pre-annealed in situ at a pressure of 1 x $10^{-5}$ Torr for 1 hr at a temperature of 400 °C. Figure 4 (f) shows an optical image of a flake prior to pre-annealing and plasma exposure for 5 min at a bias of +60 V. The flake contained MLG, BLG, tri-layer graphene (TLG) and multi-layer regions up to about $n$ = 15. After exposure, the entire flake disappeared, indicating a removal rate of at least 3 layers/min. **Figure 5** (a) shows an optical image of a flake prior to pre-annealing and plasma exposure for 1 min at a bias of +60 V. After exposure, only the MLG region was completely removed, as shown in Figure 5 (b). The sole removal of the MLG region is corroborated by the absence and presence of MLG and BLG steps in the AFM image of Figure 5 (c), respectively. For this sample, layers were removed at an average rate of 1 layer/min, 2.5 times faster than the rate for the un-annealed sample in Figure 4 (a). Thus, the rate for removing the initial 1-2 layers from pre-annealed samples is higher than that for un-annealed samples. We note that annealing of graphene has been reported to remove adsorbates from the basal plane, as inferred from electrical transport measurements. For MLG, annealing at 150 °C has been reported to desorb $H_2O$, $NH_3$, CO and $NO_2$ adsorbates, and annealing at 200 °C to desorb $O_2$ molecules.[27,28] Mechanical exfoliation of flakes onto $SiO_2$/Si substrates in air at 50% relative humidity produces flakes that have $O_2$ and $H_2O$ adsorbed on the top and bottom surfaces from contact with the air and the substrate.[29,30] We propose that such adsorbates shield the underlying graphene from electron irradiation. We have previously shown that short exposures to electron irradiation from He plasmas (for example, 1 min with the sample electrically floating) result in the reversible hydrogenation of graphenes.[20] Our proposed mechanism for hydrogenation was electron impact fragmentation of adsorbed water vapor. We propose that during the initial stage of exposure, the electron-surface interaction is dominated by the fragmentation of adsorbates–resulting in hydrogenation. Thinning begins after the threshold timescale for the fragmentation of all adsorbates has occurred–at which point, the dissociation of chemisorbed H species followed by C atoms takes place. Pre-annealed samples would be void of adsorbates, thus thinning would occur faster for such samples.

**Figure 6** (a) shows an optical image of a pristine graphene flake with a MLG region indicated by the number 1. Figures 6 (b)-(f) are optical images of the same flake after five consecutive He plasma exposures of 6, 2, 2, 2 and 3 min (totaling 6 min, 8 min, 10 min, 12 min, and 15 min, respectively) with the sample electrically floating. After 6 min of He plasma exposure, an evenly distributed reduction of optical contrast for the MLG region of the flake is observed. We attribute this reduction to the removal of C atoms. After an additional 2 min of exposure, a further reduction of optical contrast is observed, yet the topography of the MLG region is similar to that of a pristine sample, as shown by the AFM image in the inset of Figure 6 (c). As shown in Figures 6 (d) and (e), the MLG is virtually unobservable after additional 2 and 2 min exposures, respectively, yet no change in topography is observed. As shown by the optical and AFM images in Figure 6 (f), the MLG region has completely disappeared, indicating the complete removal of one atomic layer after an additional exposure time of 3 min – for a total time of 15 min, comparable to the time required for the removal of one layer of the sample shown in Figure 1. As shown in Figures 1 and 6, the thinned multi-layer samples do not exhibit any reduction in lateral dimensions or deep pits in the basal plane. These characteristics were evident in all thinned samples, even those that were thinned by more than 30 layers. We propose that the intermediate reduction of optical contrast in Figure 6 is due to the uniform removal of C atoms from the basal plane. This hypothesis is corroborated by the continuous reduction of optical contrast and instantaneous disappearance of topography for a single atomic layer. We propose that the removal of C atoms occurs as an evenly distributed top-down phenomenon that results in the atomic-scale perforation of the exposed layer. Such vacancies created by this type of removal would neither produce large etch pits nor would they be detectable by AFM as such features are smaller than the radii of AFM probes. Furthermore, such vacancies would explain the low intensity Raman D bands observed in Figure 3 (e). We note the dependence of removal rate consistency on sample bias. The MLG samples shown in Figures 1, 3 and 6 thinned while electrically floating exhibited thinning rates of 0.08, 0.03 and 0.07 layers/min, respectively. We attribute the fluctuations in removal rates for electrically floating samples to inconsistent plasma morphologies that are not observed for plasmas ignited under a sample bias. Adding to this inconsistency in morphologies, small amounts of surface charging from an inability to neutralize charging on the floating surface may lead to variable fluctuation rates.

Electrons incident at kinetic energies of about 60 eV do not have sufficient momentum to knock out strongly bonded C atoms in the basal plane that typically have binding energies of 4-5 eV.[31] However, such kinetic energies are in the range corresponding to the maximum cross-section for electron stimulated desorption of adsorbates and fragmentation of molecules.[5,6] These processes involve electronic excitation of the adsorbate or molecule by the incident electrons, followed by desorption or fragmentation. We propose that the mechanism for layer removal involves a similar process in which C atoms are excited by the incident electrons, followed by dissociation of the C atom.

3. Conclusions

We observe the removal of single layers from exfoliated multi-layer graphene using low-energy electron irradiation from a He plasma. The unique role of electron interaction rather than ion or neutral species interaction in layer removal is observed from biasing experiments. Our technique could be used as a scalable means of patterning thinned regions from multi-layer graphene samples for device applications such as transparent conducting windows. For example, a mask could be used in conjunction with e-beam lithography for the thinning of select multi-layer regions for device applications. Our technique is not observed to produce large pits in the basal plane or a reduction in lateral dimensions of the sample. We propose that layer removal is due to the dissociation of C atoms from the basal plane due to electronic excitation by the incident electrons. The rate of layer removal varied from 0.08 to at least 3 layers/min depending on exposure time, sample bias and pre-annealing treatments. We attribute the accelerated rate of layer removal for longer exposure times to the simultaneous removal of more than one layer. The desorption of O and $H_2O$–by pre-annealing treatments–is observed to increase the rate of initial layer removal. We attribute this phenomenon to the shielding of the underlying layer by adsorbates.

**4. Experimental Section**

Our samples were produced by mechanical exfoliation of ZYB-grade highly-oriented-pyrolytic graphite onto *p*-doped Si substrates having a 300 nm thermally grown oxide layer.[1] Prior to exfoliation, substrates were cleaved into 1 by 2 cm pieces. The HF etching of the oxide layer from either ends of the cleaved substrate allowed for the application of a current and potential for pre-annealing and biasing of samples, respectively. Graphene flakes having MLG regions near BLG and TLG regions were chosen as direct experimental comparison could be drawn due to the proximity of adjacent layers.

The He plasma was ignited within a quartz tube attached to the plasma chamber at a pressure of about 50 mTorr and flow of 1-10 sccm of He, corresponding to a low pressure "cold" remote plasma. He was chosen rather than heavier inert gases to avoid possible damage from heavy nuclei during plasma exposure. Two Cu electrodes at either end of the quartz tube–attached to the RF power supply–were used for plasma ignition. To avoid energetic plasma species, samples were placed approximately 15 cm from the main excitation region. In the absence of a DC sample bias, samples were electrically floating with respect to the plasma chamber walls and earth ground.

A Thermo Electron Almega XR micro-Raman spectrometer was used for the collection of all Raman data. A 532nm solid state laser operating at a power of 6 mW was focused through the 100X objective lens of the microscope creating a spot size on the sample having a diameter $d = 0.9$ μm. A spectral resolution of 2 cm$^{-1}$ was obtained by utilizing multiple gratings. Collecting spectra between 400 cm$^{-1}$ and 4000 cm$^{-1}$ for 2 s is sufficient for recording all vibrational modes of graphenes.

**Acknowledgements**

We thank Joshua Wahrmund for useful discussions. This work was supported by the Faculty Research Grant Program, Center for Advanced Research and Technology and the Laboratory of Imaging Mass Spectrometry at the University of North Texas. Dr. Verbeck's lab acknowledges support from AFOSR-YIP-FA9550-08-1-0153.

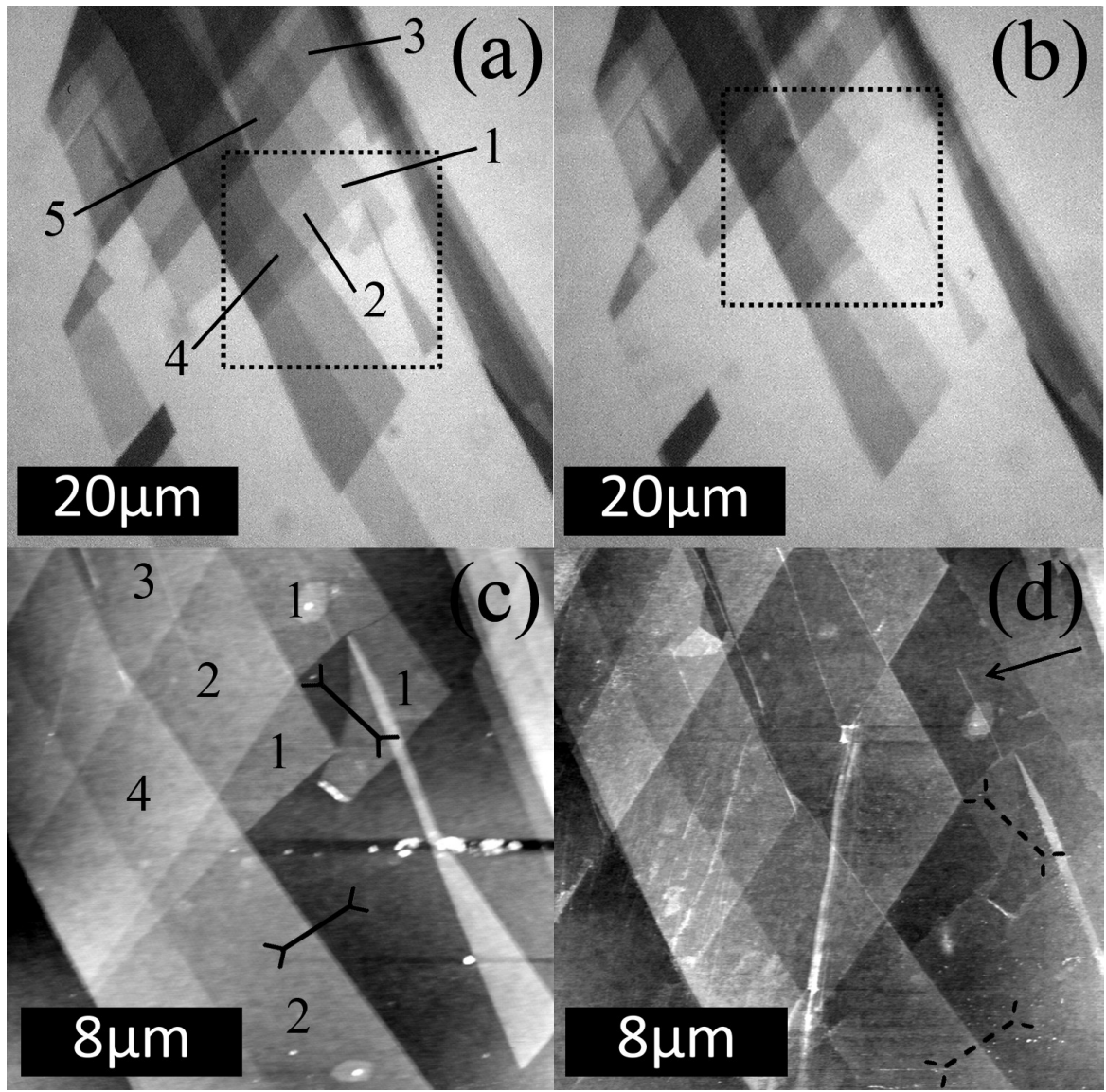

**Figure 1.** a,b) Optical images of a graphene flake before (a) and after (b) He plasma exposure. Numbers indicate the number of layers. c,d) AFM images of the boxed regions in (a,b), respectively. The solid and dashed lines indicate the locations along which step height profiles were obtained.

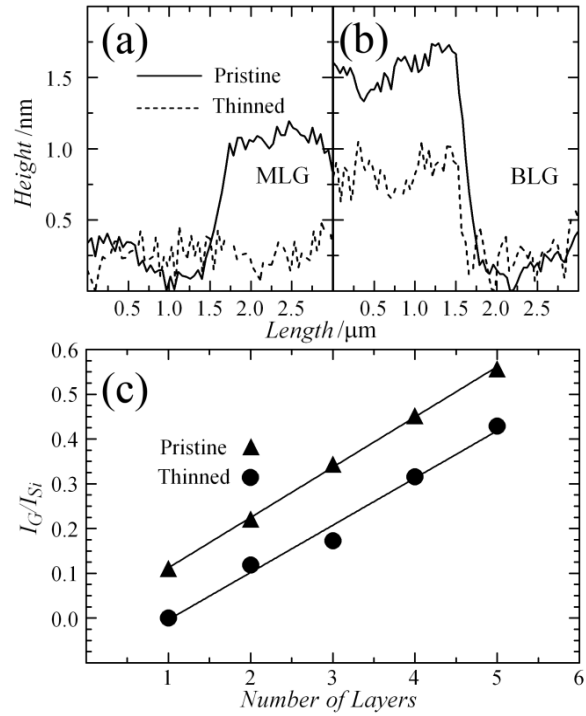

**Figure 2.** a,b) Step height profiles of the MLG (a) and BLG (b) regions, before and after He plasma exposure over the lines shown in Figure 1c,d. c) Plots of $I_G/I_{Si}$ vs $n$ before and after exposure for the regions having $n$ = 1, 2, 3, 4, 5 in Figure 1a. After exposure, the linear relationship is preserved with an overall downshift of the curve by approximately 0.13.

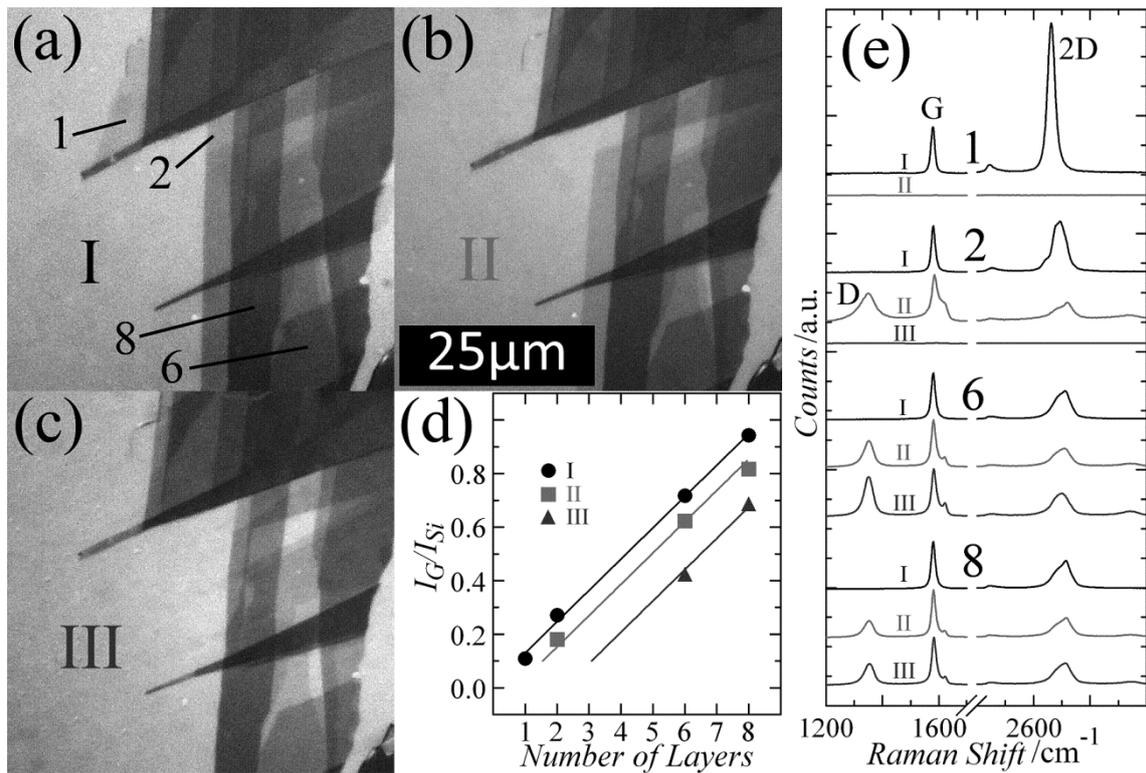

**Figure 3.** a–c) Optical images of a graphene flake before (a indicated by I) and after two consecutive He plasma exposures, indicated by II (b) and III (c). Numbers indicate the number of layers. d) Plots of $I_G/I_{Si}$ vs $n$ before and after the two consecutive exposures for the regions having $n$ = 1, 2, 6, and 8. After exposure, the linear relationship is preserved with overall downshifts of 0.09 and 0.18 for the first and second exposure, respectively. e) Raman spectra of regions indicated in (a) before and after plasma exposures.

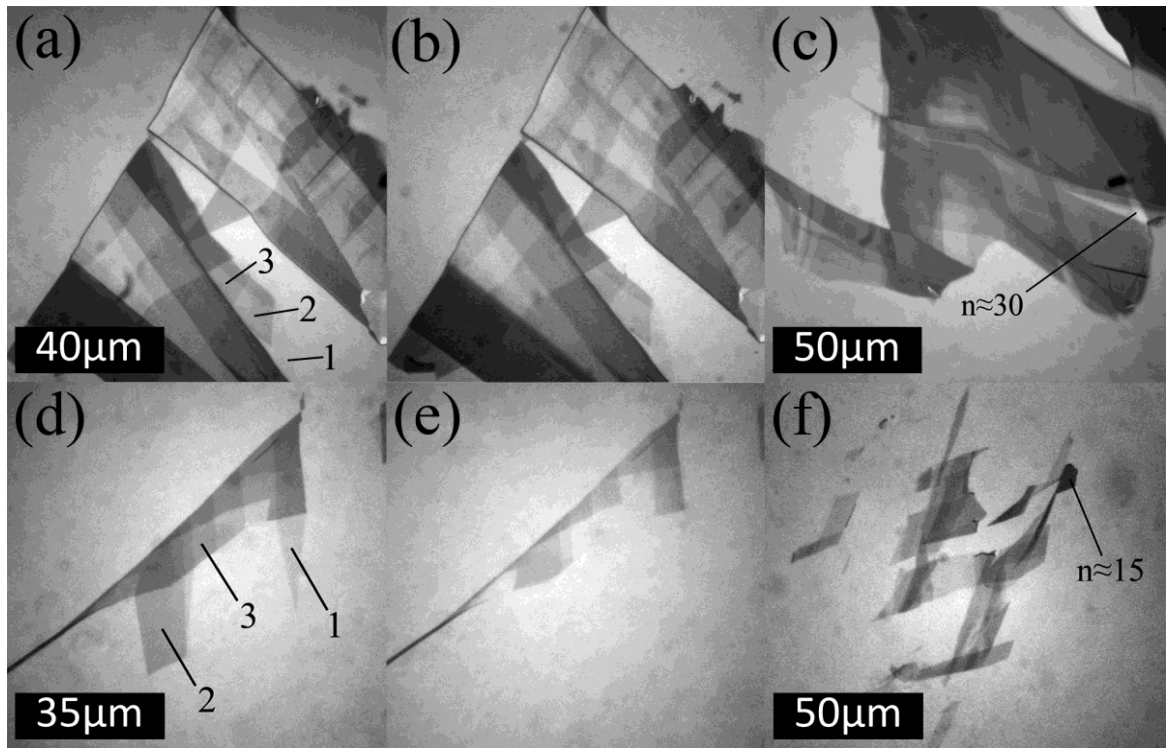

**Figure 4.** a,b) Optical images of a graphene flake before (a) and after (b) He plasma exposure, at a sample bias of -60 V. Numbers indicate the number of layers. No change in optical contrast is observable for any region. c) Optical image of a graphene flake prior to He plasma exposure at a bias of +60 V for 30 min. After the exposure, all layers of the flake were removed. d) Optical image of a graphene flake before exposure. Numbers indicate the number of layers. e) Sample shown in (d) after He plasma exposure at a bias of +60 V for 5 min. The MLG and BLG regions have been removed while the remaining TLG region is still visible, indicating the removal of about two layers. f) Optical image of a graphene flake that was pre-annealed at 400 °C for 1 hr prior to He plasma exposure at a bias of +60 V for 5 min. After the exposure, all layers of the flake were removed.

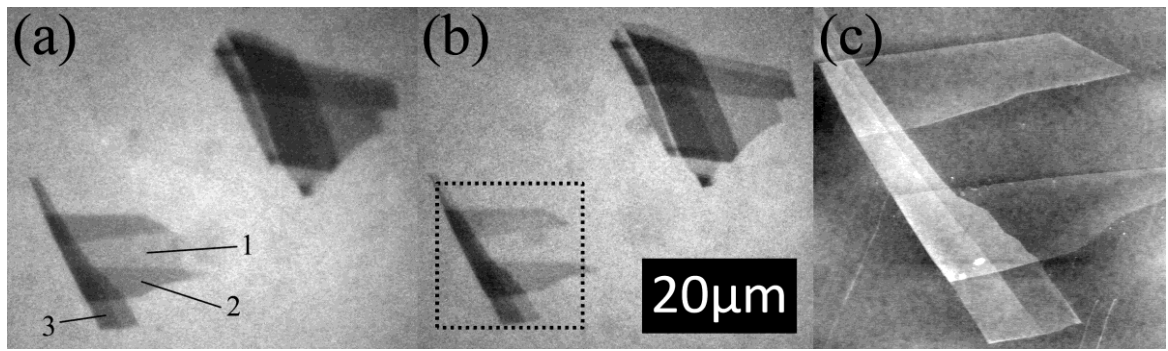

**Figure 5.** a) Optical image of a graphene flake. Numbers indicate the number of layers. b) Sample shown in (a) after pre-annealing at 400 °C for 1 hr and then exposing to He plasma at a bias of +60 V for 1 min. c) AFM image of the dashed square in (b). The MLG region is no longer visible, indicating the removal of one layer.

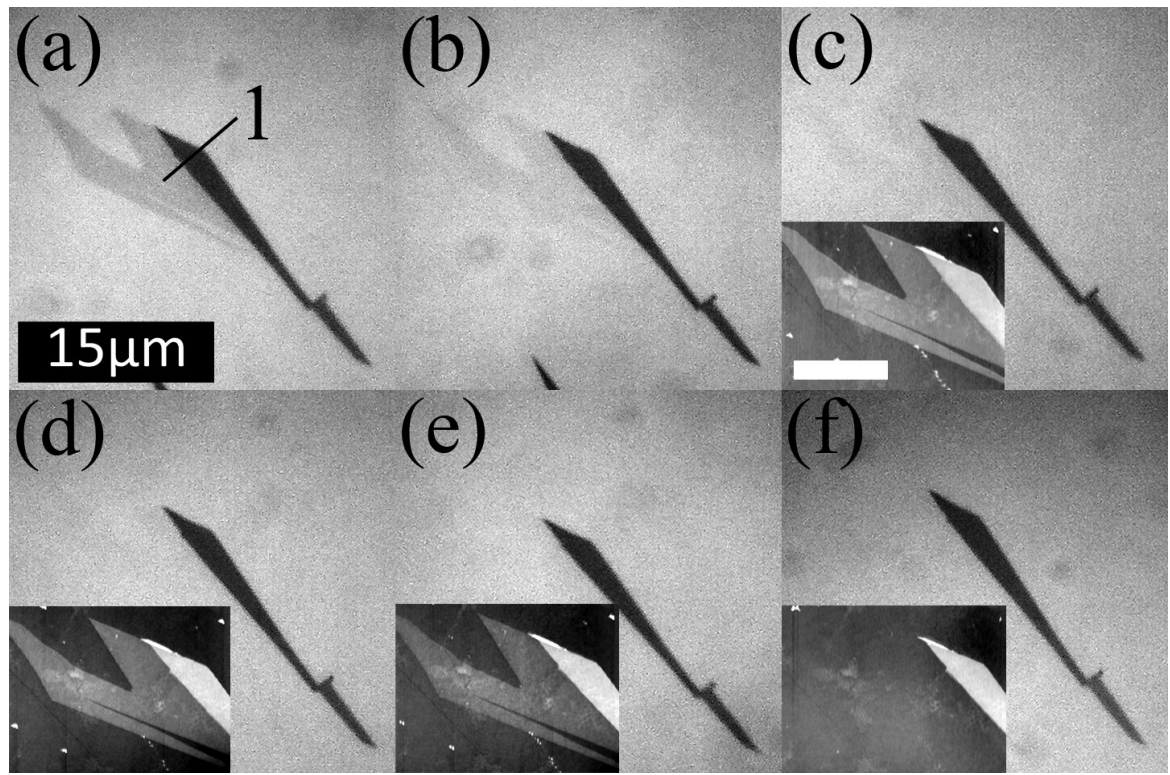

**Figure 6.** a) Optical image of a pristine graphene flake. The MLG region is indicated by the number 1. b–f) Optical images of the same sample after 6 (b), 8 (c), 10 (d), 12 (e), and 15 (f) min of He plasma exposure. Insets in c–f show complimentary AFM images. Scale bar for insets equals 5 μm.

**The removal of single atomic layers from multi-layer graphene** by electron bombardment from a cold He plasma is reported. We propose that the evenly distributed removal of carbon atoms occurs due to electron excitation followed by the dissociation of atoms from the graphene basal plane. This simple technique could be used as a novel approach for the patterned fabrication of thin film transparent conducting electrodes from multi-layer graphenes.

Thinning

J. D. Jones, R. K. Shah, G. F. Verbeck, and J. M. Perez

ToC figure ((55 mm broad, 40 mm high))

Page Headings
Left page:     J. D. Jones et al.
Right page:    Thinning of Multi-Layer Graphene